\documentclass[article,aps,twocolumn,showpacs,amsmath,amssymb,superscriptaddress,nofootinbib,floatfix]{revtex4-2}

\usepackage{graphicx}
\usepackage{dcolumn}
\usepackage{bm}
\usepackage[colorlinks=true,linkcolor=blue,citecolor=blue,urlcolor=blue]{hyperref}

\usepackage{array,booktabs,tabularx}
\usepackage[caption=false]{subfig}
\usepackage[USenglish]{babel}
\usepackage{braket}
\usepackage{xcolor}
\usepackage{amsfonts}
\usepackage{amssymb}
\usepackage[normalem]{ulem}
\usepackage{amsmath}

\begin{document}

\title{Flat band driven itinerant magnetism in the Co-pnictides CaCo$_2$As$_2$ and LaCo$_2$P$_2$}

\author{D. Subires}
\email{david.subires@dipc.org}
\affiliation{Donostia International Physics Center (DIPC), Paseo Manuel de Lardizábal. 20018, San Sebastián, Spain}
\affiliation{Applied Physics Department, University of the Basque Country (UPV/EHU), Basque Country, Bilbao, 48080 Spain}

\author{M. García-Díez}
\affiliation{Donostia International Physics Center (DIPC), Paseo Manuel de Lardizábal. 20018, San Sebastián, Spain}
\affiliation{Physics Department, University of the Basque Country (UPV/EHU), Leioa, Spain}

\author{A. Kar}
\affiliation{Donostia International Physics Center (DIPC), Paseo Manuel de Lardizábal. 20018, San Sebastián, Spain}

 \author{C.-Y. Lim}
\affiliation{Donostia International Physics Center (DIPC), Paseo Manuel de Lardizábal. 20018, San Sebastián, Spain}

\author{Victoria M. Li}
\affiliation{Department of Chemistry and
Biochemistry, Florida State University, Tallahassee, Florida
32306, United States}

\author{V Yannello}
\affiliation{Department of Chemistry and
Biochemistry, University of Tampa, 401 W. Kennedy Blvd. Tampa, FL 33606}

\author{Dina Carbone}
\affiliation{MAX IV Laboratory, Fotongatan 2, Lund, 225 92, Sweden}

\author{P. Gargiani}
\affiliation{ALBA Synchrotron Light Source, Cerdanyola del Vall\`es, 08290 Barcelona, Catalonia, Spain}

\author{T. Yilmaz}
\affiliation{National Synchrotron Light Source II, Brookhaven National Laboratory, Upton, New York 11973, USA}
\affiliation{Department of Physics, Xiamen University Malaysia, Sepang 43900, Malaysia}

\author{J. Dai}
\affiliation{ALBA Synchrotron Light Source, 08290 Barcelona, Spain}

\author{M. Tallarida}
\affiliation{ALBA Synchrotron Light Source, 08290 Barcelona, Spain}

\author{E. Vescovo}
\affiliation{National Synchrotron Light Source II, Brookhaven National Laboratory, Upton, New York 11973, USA}

\author{M. Shatruk}
\affiliation{Department of Chemistry and
Biochemistry, Florida State University, Tallahassee, Florida
32306, United States}

\author{Maia G. Vergniory}
\affiliation{Donostia International Physics Center (DIPC), Paseo Manuel de Lardizábal. 20018, San Sebastián, Spain}
\affiliation{Département de Physique et Institut Quantique,
Université de Sherbrooke, Sherbrooke, J1K 2R1 Québec, Canada.}

\author{S. Blanco-Canosa}
\email{sblanco@dipc.org}
\affiliation{Donostia International Physics Center (DIPC), Paseo Manuel de Lardizábal. 20018, San Sebastián, Spain}
\affiliation{IKERBASQUE, Basque Foundation for Science, 48013 Bilbao, Spain}

\begin{abstract}
Flat bands can induce strong electron correlation effects that help stabilize both magnetic and superconducting states. Here, we carry out angle-resolved photoemission spectroscopy and density functional theory calculations to study the electronic structure of the Co-pnictides CaCo$_2$As$_2$ and LaCo$_2$P$_2$. We find that the \textit{k}$_z$ Fermi topology of ferromagnetic LaCo$_2$P$_2$ is markedly 2-dimensional, while the antiferromagnetic CaCo$_2$As$_2$ develops a 3D Fermi surface described by a \textit{zig-zag}-like band dispersion perpendicular to the Co-As plane. Furthermore, the magnetism is driven by the electronic correlations of the flat bands with \textit{d}$_{xy}$ and \textit{d}$_{z^2}$ orbital character at the Fermi level. Our results link the electronic dimensionality and the magnetic order, and emphasize the critical role of the As-As and P-P bond strength along the \textit{c}-direction to understand the electronic band structure and the rich phase diagram of transition metal pnictides.
\end{abstract}

\maketitle

The unique interplay between electron-electron correlations and the high density of states at the Fermi level results in strong interactions that throw off their balance and favor unconventional orders. In the case of weakly dispersive flat bands, Coulomb repulsion dominates over the kinetic energy, giving rise to enhanced correlations especially in 2-dimensional (2D) materials and kagome metals that favor magnetism, fractional quantum Hall states and competing phases such as superconductivity, enhanced \textit{d}$^0$ magnetism and charge density waves \cite{Wang_2023,Yin_2019,Ren_2024,Qian_2014,Brit_2011,Jonah_2022,Lin_2018,Kang_2020,Cao_2024}.

Transition metal pnictides with a tetragonal structure, described by the chemical formula \textit{A}T$_2$X$_2$ (A: alkali metal, alkaline-earth metal, lanthanide; T: transition metal; X: metalloid), have garnered renewed attention since the discovery of superconductivity, charge/spin density waves (CDW/SDW) and nematicity in the Fe-based family, \textit{A}Fe$_2$As$_2$ \cite{Fernandes2022,Si2016}, and broken rotational and translational symmetries in the Ni-based pnictides \cite{Yao2022,Song2022}. Notably, the Co-pnictides, \textit{A}Co$_2$X$_2$, do not develop either CDW or superconducting phases at low temperature, but an intriguing cascade of magnetic phase transitions strongly linked to their electronic structure \cite{Reehius1998}. Besides, the substitution of the A ions [A= Ca, Sr, La, Nd, Pr, Eu,...] in \textit{A}Co$_2$X$_2$ \cite{Imai2015} is accompanied by a variation of the Co$_2$X$_2$ interlayer distance and X-X bond angle, leading to a collapsed tetragonal (\textit{c}T) structure \cite{Tan2016,Tan2018} that modify their electronic structure and magnetic order \cite{Clark2020,Jia2009}. 

Although the Co$_2$X$_2$ plane is intrinsically 2D, the dimensionality and magnetic order of Co-pnictides can be tuned by electron counting and pressure. For instance, LaCo$_2$P$_2$, with an interlayer P-P distance of 3.16 \r{A}, is an in-plane polarized ferromagnet (FM), with T$_\mathrm{C}$$\simeq$130 K \cite{Reehius1994,Zhang_2015}, while CaCo$_2$As$_2$ (As-As distance of $\simeq$2.73 \r{A} has been reported to develop an A-type antiferromagnetic (AFM) ground state (T$_\mathrm{N}$=76 K) coexisting with FM spin fluctuations within the CoAs layer \cite{Cheng_2012} and a spatial anisotropy due to the magnetic frustration arising from the competing FM and AFM interactions \cite{Sapkota2017}. Magnetic order has been observed in the Sr(Ni$_{1-x}$Co$_x$)$_2$P$_2$ series \cite{Schmidt_2023}, underscoring the effect of the \textit{c}/\textit{a} ratio and chemical doping on the magnetic and electronic properties of Co-pnictides. Similarly, the cascade of magnetic phase transitions in the (Ca,Sr)Co$_2$P$_2$ series is accompanied by a change from itinerant-electron FM to AFM order \cite{Jia2009,Sugiyama2015,Ying_2012}, which uncovers novel forms of collective ground states, such as Weyl fermions \cite{Xu2020} and Kondo physics \cite{Poelchen_2022}. Indeed, CeCo$_2$P$_2$ has been reported to be a topological Kondo lattice compound driven by the flat band induced strong interactions and magnetism \cite{Liu_2024,Haoyu2024}. 

On the other hand, despite the magnetism reported experimentally in Co-pnictides, the link between magnetic correlations, chemical structure, and electronic dimensionality/band structure has not been hitherto addressed. The microscopic origin of the magnetic orders is particularly appealing, since Density Functional Theory (DFT) calculations unveil the presence of a flat band along the $\Gamma$-M direction and high density of states (DOS) at the Fermi level, which potentially would give rise to magnetic order through the Stoner mechanism. This debate is further stimulated by the observation of flat band induced magnetism in the sibling transition metal dichalcogenide \textit{A}Co$_2$Se$_2$ compound \cite{Yang_2013,Huang_2021} and the coexistence of both AFM and FM spin fluctuations in SrCo$_2$As$_2$ \cite{Li_2019}.  

Here, we use angle-resolved photoemission spectroscopy (ARPES) and DFT calculations to link the chemical structure, Fermi surface topology, flat bands, and magnetic order in ACo$_2$X$_2$ (A= La, Ca, and X = P, AS). We find that the electronic structure of LaCo$_2$P$_2$ is 2D, in contrast to the 3D observed in the itinerant antiferromagnet CaCo$_2$As$_2$. The 3D Fermi surface topology in CaCo$_2$As$_2$ favors the interlayer antiferromagnetic order, provided that the As-As distance, \textit{d}$_\mathrm{As-As}$, is shorter than \textit{d}$_\mathrm{P-P}$ in LaCo$_2$P$_2$. From the DFT calculations, we achieve an accurate description of the orbital contribution to the band structure and demonstrate that the magnetic moment is dominated by the \textit{d}$_{xy}$ and \textit{d}$_{z^2}$ orbitals of Co, that form a flat band along the $\Gamma$-$\mathrm{M}$ direction. Our work brings an important piece of information to understand the intertwining of electronic dimensionality and the magnetic order in the family of layered AT$_2$X$_2$-type compounds
with flat band driven magnetism.

\begin{figure}
    \centering
    \includegraphics[width=\columnwidth]{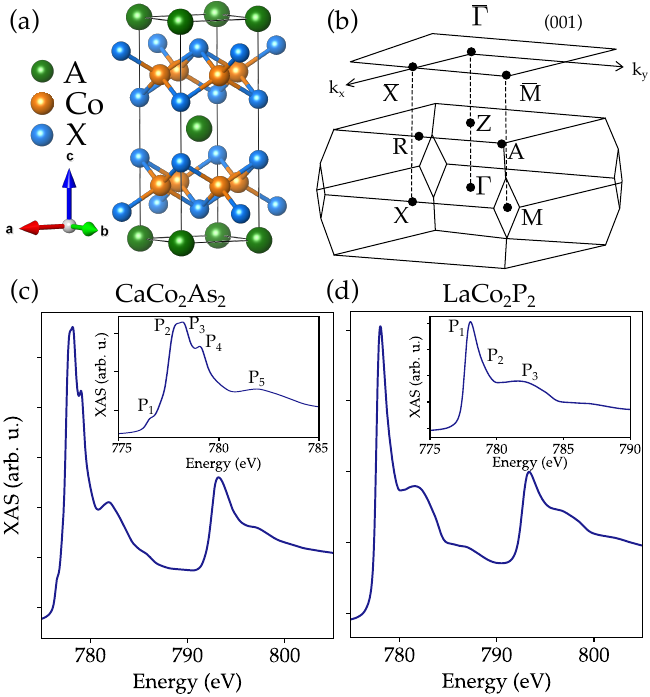}
    \caption{(a) Conventional unit cell of the Co-pnictides with $\mathrm{ThCr_2Si_2}$-type structure, where A = Ca, La, and X = As, P. (b) Brillouin zone (BZ) and the projected BZ on the (001) surface, along the high symmetry directions. (c-d) Isotropic ($\sigma^+$+$\sigma^-$) x-ray absorption spectroscopy (XAS) of $\mathrm{CaCo_2As_2}$ and $\mathrm{LaCo_2P_2}$, respectively, labeling the different transitions in the inset.}
    \label{figure1}
\end{figure}

Single crystals of CaCo$_2$As$_2$ and LaCo$_2$P$_2$ were grown by Sn-flux method, as previously reported elsewhere \cite{Shatruk2019,Tan2016,Tan2018,Mann2022}. x-ray absorption spectroscopy (XAS) and x-ray magnetic circular dichroism (XMCD) experiments at the Co $\mathrm{L_{2,\,3}}$-edge were carried out at the BOREAS beamline at the ALBA synchrotron. ARPES experiments were performed at the BLOCH (MAX IV), ESM (NSLS-II), and LOREA (ALBA) beamlines, respectively, with experimental energy and angular resolution of $10\,\mathrm{meV}$ and $0.1^\circ$. The samples were in-situ cleaved and the measurements were taken at 20 K and above the transition temperatures. DFT calculations of the electronic band structure were carried out using the \textit{Vienna Ab-initio Simulation Package} (VASP) software\cite{vasp1,vasp2} for the experimental structures available on Materials Project \cite{materialsproject}. We chose the Perdew-Burke-Ernzenhof (PBE) implementation of the Generalized Gradient Approximation (GGA) for the exchange-correlation functional \cite{gga_pbe} together with Projector Augmented Wave (PAW) pseudo-potentials \cite{paw}. The $k$-point grid was adjusted to a Gamma-centered, regular Monkhorst-Pack point set of dimensions $13\times 13\times 13$, and the plane wave basis energy-cutoff was fixed to 600 eV. To accurately describe the \textit{d} valence electrons, a series of DFT+U calculations in the rotational-invariant implementation \cite{dftu} were performed to fit the optimal value of the repulsive \textit{U} parameter in the \textit{d} subspace to reproduce the reported experimental magnetic moment \cite{Cheng_2012,laco2p2_magnetic_moment}. It was determined that \textit{U}=1 for AFM $\mathrm{CaCo_2As_2}$ and \textit{U}=1 for FM $\mathrm{LaCo_2P_2}$, with resulting magnetic moments of $0.6\mu_B$ and $0.485\mu_B$ per atom, respectively. A Wannier function basis accurately describing the band structure within $\pm 1$ eV from the Fermi level was constructed using Wannier90 \cite{wannier90}. Furthermore, we used the interpolated tight-binding Hamiltonian obtained in this basis to compute the Fermi surface and surface states using WannierTools \cite{wanniertools,SI}.

The Co-pnictides CaCo$_2$As$_2$ and LaCo$_2$P$_2$ exhibit a body-centered tetragonal ThCr$_2$Si$_2$ structure (space group $I4/mmm$, no. 139) characterized by alternating stacks of $\mathrm{CoX_2}$ tetrahedral layers (Fig. \ref{figure1}(a)), with the strong bonding within the $\mathrm{CoX_2}$ tetrahedra dominating the interactions, akin to the chemical structure of Fe-pnictides \cite{Ozawa_2008}. The conventional unit cell lattice parameters are \textit{a} = \textit{b} = 4.01 \r{A} and $c_c$= 10.38 \r{A} for CaCo$_2$As$_2$ and \textit{a}= \textit{b} = 3.80 \r{A} and $c_c$= 11.01 \r{A} for LaCo$_2$P$_2$, with the ratio $c_c$/\textit{a} = 2.59 and 2.9, respectively, suggesting a structural link with the magnetic order \cite{Tan2018}. 

\begin{figure*}
    \centering
    \includegraphics[width=\linewidth]{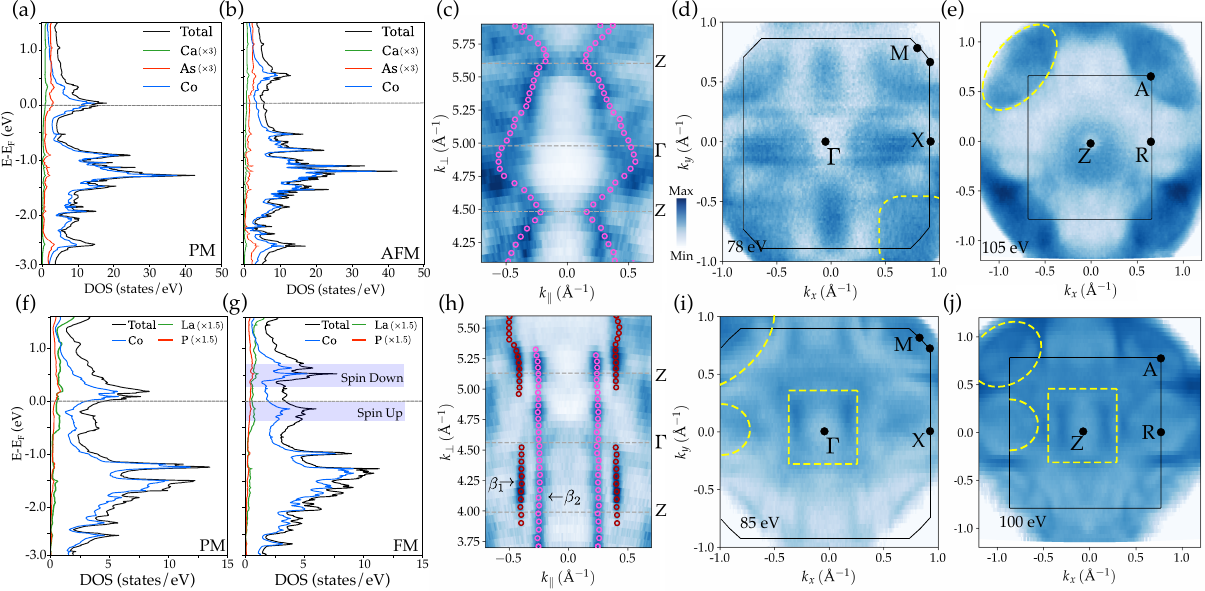}
    \caption{(a-b) Atomically resolved density of states (DOS) of $\mathrm{CaCo_2As_2}$ for the PM (a) and AFM (b) phases. (c) \textit{k}$_x$-\textit{k}$_z$ Fermi surface of $\mathrm{CaCo_2As_2}$ at T=20 K exhibiting a \textit{zig-zag} dispersion along the \textit{c}-direction, superimposed with the Fermi momentum, (\textit{k}$_f$), extracted from the momentum distribution curves (MDCs). (d,e) Fermi surface of $\mathrm{CaCo_2As_2}$, corresponding to the $k_z=0$ and $k_z=\frac{\pi}{c}$ planes, taken with $h\nu=78\,\mathrm{eV}$ and $h\nu=105\,\mathrm{eV}$, respectively. (f-g) Total and atomically resolved DOS for the PM (f) and FM (g) order of $\mathrm{LaCo_2P_2}$. (h) \textit{k}$_x$-\textit{k}$_z$ Fermi surface of $\mathrm{LaCo_2P_2}$, T=20 K. (i,j) \textit{k}$_x$-\textit{k}$_y$ Fermi surface at the $k_z=0$ and $k_z=\frac{\pi}{c}$ planes of $\mathrm{LaCo_2P_2}$, taken with $h\nu=85\,\mathrm{eV}$ and $h\nu=100\,\mathrm{eV}$, respectively, and T=20 K. The yellow dashed lines that highlight the electron pockets are a guide to the eye.}
    \label{figure2}
\end{figure*}

Figs. \ref{figure1} (c-d) displays the X-ray absorption (XAS) spectra corresponding to the Co \textit{L}$_3$ ($\sim780$ eV) transition that already hints at a different electronic structure. The XAS lineshape strongly depends on the multiplet structure, mainly given by the Co 3\textit{d}$\rightarrow$3\textit{d} and 2\textit{p}$\rightarrow$3\textit{d} Coulomb and exchange interactions, the local crystal fields, and the hybridization with the ligands \cite{Haverkort_2012}.
The absorption edges are separated by $\sim$16 eV from the \textit{L}$_2$ edge ($\sim795$ eV) that results from the spin-orbit split 2\textit{p}$\rightarrow$3\textit{d} transitions. The Co \textit{L}$_3$ spectrum of CaCo$_2$As$_2$ presents five fine structures: a `pre-edge' transition at 776 eV (P$_1$), a major absorption edge with 2 main peaks at 777.8 eV (P$_2$) and 778.2 eV (P$_3$), two high energy shoulders at 779 eV (P$_4$) and 782 eV (P$_5$), characteristic of a Co$^{2+}$ oxidation state, see Fig. \ref{figure1} (c), and a broad absorption line corresponding to the transitions into the continuum above 786 eV. On the other hand, the \textit{L}$_3$ spectrum of LaCo$_2$P$_2$ is composed of a main peak (P$_1$ at 778.4 eV) and two shoulders at 779 eV (P$_2$) and 782.5 eV (P$_3$), Fig. \ref{figure1} (d), suggesting the continuum of transitions 10 eV above the main edge \cite{Haverkort_2006,Ahad2020}.

The calculated DOS of CaCo$_2$As$_2$ shows a continuum of metallic states, see Fig. \ref{figure2}. In the paramagnetic (PM) calculation, we observe signatures of hybridization between Co and As 4\textit{p} states, which contributes nearly $\simeq$8\% to the DOS of CaCo$_2$As$_2$, and the Ca participates with almost $\simeq$3\% of states in the vicinity of the Fermi level, hence, the total DOS is mainly derived from the Co 3\textit{d} orbitals that develop a sharp peak at E$_\mathrm{F}$, Fig. \ref{figure2}(a). Interestingly, the sharp peak of the DOS in the PM calculation is split $\pm$0.5 eV around the Fermi level (Fig. \ref{figure2}(b)), suggesting that electron correlations are responsible for the magnetic order. Similarly, the DOS of LaCo$_2$P$_2$, Fig. \ref{figure2}(f), presents a sharp peak near the Fermi level ($\sim5$ states/eV) in the PM phase, that splits into spin-polarized minority and majority bands in the FM phase due to the Stoner mechanism, Fig. \ref{figure2}(g) \cite{kovnir2010tuning}. 

In Figs. \ref{figure2}(c) and (h), we show the \textit{k}$_x$-\textit{k}$_z$ Fermi surface of CaCo$_2$As$_2$ and LaCo$_2$P$_2$, obtained by varying the incident photon energy from 50 to 120 eV ($\Delta$E=2 eV). The high symmetry points of CaCo$_2$As$_2$, labeled as $\Gamma$ and $\mathrm{Z}$, are located at 78 and 105 eV, respectively, Fig. \ref{figure2}(d-e). The \textit{zig-zag} dispersion of the bands, superimposed with the Fermi momenta (\textit{k}$_f$) obtained from the momentum distribution curves (MDC) of the bands crossing the Fermi level along the \textit{c}-direction, demonstrates a 3D character of its Fermi surface of CaCo$_2$As$_2$, Fig. \ref{figure2}(c). On the other hand, the \textit{k}$_z$ dependence of LaCo$_2$P$_2$ displays two dispersionless bands, identified as $\beta$$_1$ and $\beta$$_2$ in Fig. \ref{figure2}(h), revealing a pronounced 2-dimensionality of its electronic structure, consistent with the high magnetic field transport data reported in the literature \cite{Teruya2014}.

Figures \ref{figure2}(d-e) and \ref{figure2}(i-j) display the low temperature \textit{k}$_x$-\textit{k}$_y$ Fermi surface of CaCo$_2$As$_2$ and LaCo$_2$P$_2$ at \textit{k}$_z$=0 and $k_z=\frac{\pi}{c}$ planes, being $c$ the lattice parameter of the primitive unit cell (distance between two Co planes), corresponding to the $\Gamma$ and $\mathrm{Z}$ planes. The constant energy contours of the electronic structure present \textit{C}$_4$ rotational symmetry, with a rectangular electron pocket at the BZ center \cite{PRB_Mansart_ARPES,PRB_Yin} and 4 square electron pockets at the corners, $\mathrm{M}$ and A points, as expected from the $I4/mmm$ crystal space group.

\begin{figure*}
    \centering
    \includegraphics[width=\linewidth]{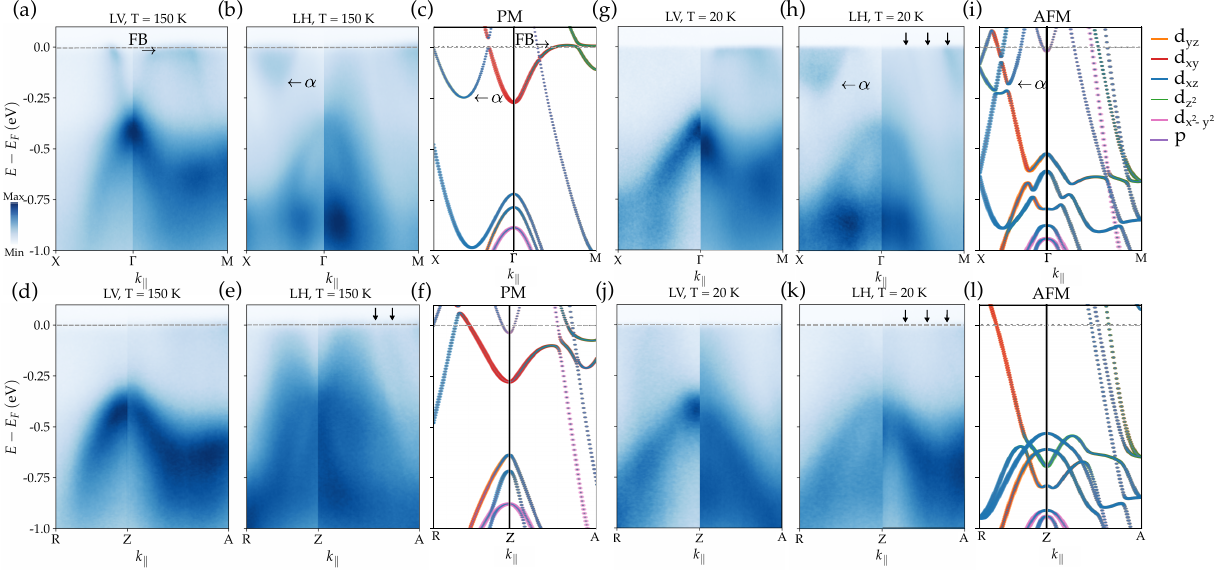}
    \caption{(a-b) Valence band dispersion of CaCo$_2$As$_2$ along the high symmetry $\mathrm{X}-\Gamma-\mathrm{M}$ line at $k_z=0$ (E$_\mathrm{i}$=78 eV), T=150 K. (c) Orbital resolved DFT calculation of the \textit{d} orbitals of Co at $k_z=0$ in the paramagnetic state. (d,e) VB spectrum along the high symmetry $\mathrm{R}-\mathrm{Z}-\mathrm{A}$ line. (f) Atomic-orbital contribution DFT calculation of the \textit{d} orbitals of Co at $k_z=\frac{\pi}{c}$. (g-h) VB dispersion at low temperature and $k_z=0$ (i) DFT calculation for $k_z=0$, and (j-i) for $k_z=\frac{\pi}{c}$. See supp. inf. Fig. 6 for high-resolution VBs that identify the bands crossing the Fermi level in (h) and (k).}
    \label{figure3}
\end{figure*}
 
 \begin{figure*}
    \centering
    \includegraphics[width=\linewidth]{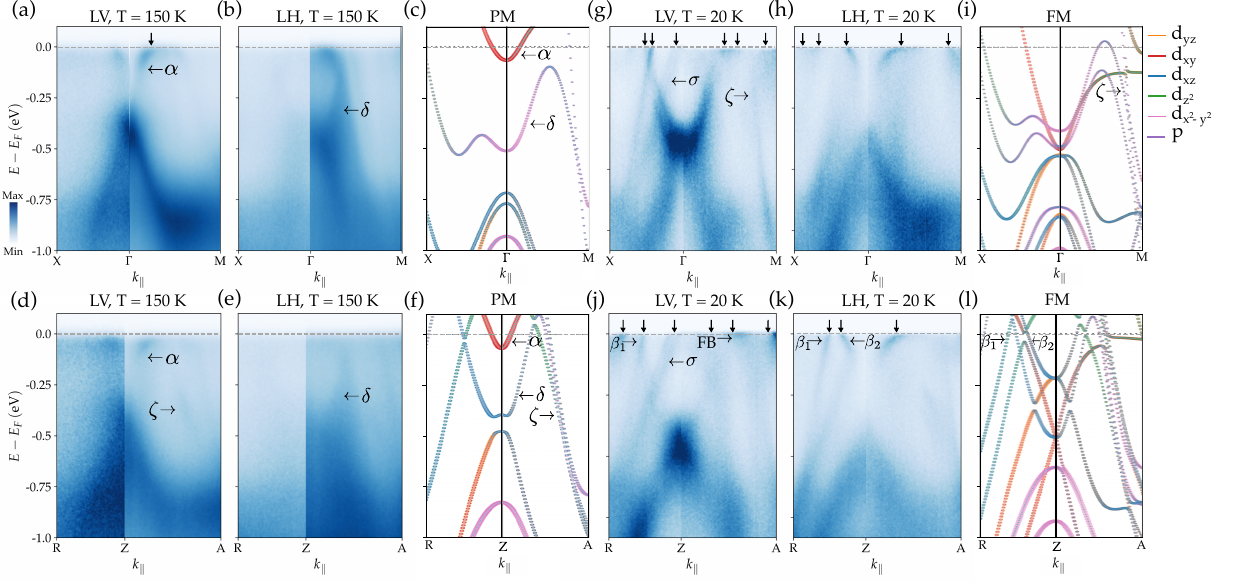}
    \caption{(a-b) VB dispersion along the high symmetry $\mathrm{X}-\Gamma-\mathrm{M}$ line for LaCo$_2$P$_2$, $k_z=0$ (E$_\mathrm{i}$=85 eV), T=150 K. (c) Orbital resolved VB at \textit{k}$_z$=0. (d,e) Band dispersion following the $\mathrm{R}-\mathrm{Z}-\mathrm{A}$ path, E$_\mathrm{i}$=100 eV. (f) Orbital contribution to the band structure at $k_z=\frac{\pi}{c}$. (g-h) Band dispersion at low temperature, T=20 K and \textit{k}$_z$=0 plane. (i) VB DFT calculation at \textit{k}$_z$=0. (j-k) Band dispersion and (l) DFT calculation at the $k_z=\frac{\pi}{c}$ plane.}
    \label{figure4}
\end{figure*}
 
Centering our attention on the \textit{k}$_z$=0 plane of CaCo$_2$As$_2$, Fig.~\ref{figure3} (a-c) displays the electronic energy-momentum band dispersion along the high-symmetry path X-$\Gamma$-M and its comparison with the projected orbital character of Co obtained \textit{ab initio} in the paramagnetic state (150 K). At $\Gamma$, there is an electron pocket derived from the \textit{d}$_{xy}$ orbital that develops more spectral weight with LV polarized light. Along the $\mathrm{X}-\Gamma$ path, we observe another electron pocket  ($\alpha$) with \textit{d}$_{xz}$ character, which is only visible with LH polarized light. In addition to these two electron pockets, our high-temperature valence band (VB) spectra resolve a flat band (FB in Fig. \ref{figure3}(a) and suppl. inf. Fig. S7) along the $\Gamma$-M direction, resulting from the hybridization of \textit{d}$_{xy}$ and \textit{d}$_{z^2}$ orbitals. The momentum spread of the flat band is responsible for the high DOS at E$_\mathrm{F}$ in Fig. \ref{figure2}(a) and contributes 0.39 $\mathrm{\mu_B}$ (\textit{d}$_{xy}$) and 0.12 $\mathrm{\mu_B}$ (\textit{d}$_{z^2}$) to the total magnetization, as calculated \textit{ab initio}. In the $k_z=\frac{\pi}{c}$ plane, there is an electron pocket centered at $\mathrm{Z}$ with $d_{xy}$ orbital character, visible with LH light, Fig. \ref{figure3}(e). Along the high-symmetry $\mathrm{Z}-\mathrm{A}$ line, we can further resolve 2 highly-dispersive bands with \textit{d}$_{x^2-y^2}$ and \textit{d}$_{xz}$ orbital characters that cross E$_\mathrm{F}$ at $k_{f}=0.4\,$\r{A}$^{-1}$ and 0.65\r{A}$^{-1}$, highlighted with vertical arrows in Fig. \ref{figure3}(e). In addition, the small electron pocket at A is also well reproduced by our DFT calculations. Nevertheless, despite the good agreement between ARPES and DFT, we note a mismatch of 0.2 eV of the broad `$\wedge$' band at $\simeq$0.4 eV below the E$_\mathrm{F}$, mostly identified with the \textit{d}$_{xz}$, \textit{d}$_{yz}$ and \textit{d}$_{x^2-y^2}$ orbitals in both \textit{k}$_z$=0 and $k_z=\frac{\pi}{c}$ planes. We attribute this discrepancy to the magnetic frustration of CaCo$_2$As$_2$ \cite{Sapkota2017} that is not fully captured by our DFT calculations. Indeed, the position of the `$\wedge$' band is well reproduced by the FM calculation, see suppl. inf. Fig. S4 further supports the frustrated square lattice scenario.  
At low temperature, Fig. \ref{figure3}(g-h), the band structure of CaCo$_2$As$_2$ is substantially modified. Notably, the electron pocket at $\Gamma$ disappears, an electron pocket is visible along the $\Gamma$-X path, Fig. \ref{figure3}(h) (\textit{d}$_{xy}$ and \textit{d}$_{xz}$ hybridized bands). Besides, along the $\Gamma-\mathrm{M}$ direction, we identify three highly-dispersive bands with Fermi momenta, $k_{f}=0.3\,$\r{A}$^{-1}$, 0.6 \r{A}$^{-1}$ and 0.8 \r{A}$^{-1}$, marked by arrows in Fig. \ref{figure3}(h) (suppl. inf. Fig. S7, for higher resolution VBs), also well reproduced by the AFM DFT calculations in Fig. \ref{figure3}(i). In the $k_z=\frac{\pi}{c}$ plane, the bands along the $\mathrm{Z}-\mathrm{A}$ path have Fermi momenta $k_{f}=0.25\,$\r{A}$^{-1}$, 0.6 \r{A}$^{-1}$ and 0.8 \r{A}$^{-1}$, similar to the highly-dispersive bands in the $k_z=0$ plane. These highly-disperse bands and the `$\wedge$'-shaped set of bands 0.4 eV below E$_\mathrm{F}$ retain their orbital characters with respect to the high temperature paramagnetic phase, Figs. \ref{figure3}(g-h) and (j-k).

The experimental band structure of the PM (150 K) and FM (20 K) phases of LaCo$_2$P$_2$ and the \textit{ab-initio} calculations are plotted in Fig. \ref{figure4}. At high temperature, the `$\wedge$' shaped band shows up at $\Gamma$ around $\simeq$0.4 eV below E$_\mathrm{F}$, Fig. \ref{figure4}(a), and its orbital contributions, \textit{d}$_{xz}$, and \textit{d}$_{x^2-y^2}$, are well discerned in the polarization dependent VB spectra. The top of this band approaches the electron pocket ($\alpha$) and crosses E$_\mathrm{F}$ at $k_{f}=0.18\,$\r{A}$^{-1}$, vertical arrow in Fig. \ref{figure4}(a), with most of its spectral weight derived from the \textit{d}$_{xy}$ orbital. In addition, the VBs with LH light show an electron-like band ($\delta$), $d_{xz}$ and $d_{x^2-y^2}$ orbital character, centered at $\Gamma$ with its bottom at a binding energy of $\sim 0.4\,\mathrm{eV}$. In the $k_z=\frac{\pi}{c}$ plane, Fig. \ref{figure4}(d-e), the electronic energy bands show similar features as in the $\Gamma-$plane, with the faint electron pocket ($\alpha$) and electron-like band ($\delta$) centered at $\Gamma$. Moreover, two highly linear dispersive bands ($\zeta$), visible for LV polarized light, are seen along the $\mathrm{Z}-\mathrm{A}$ direction, in agreement with the DFT calculations. At low temperature, Fig. \ref{figure4}(g-l), we resolve a large electron pocket around $\Gamma$ and \textit{k}$_z$$\sim$0, Fig. \ref{figure4}(g), which presents mostly \textit{d}$_{x^2-y^2}$ and \textit{d}$_{yz}$ character. Such electron pocket cuts E$_\mathrm{F}$ at $k_{f}=0.47\,$\r{A}$^{-1}$ along $\mathrm{X}-\Gamma$ and $\Gamma-\mathrm{M}$ paths, with the dispersive band ($\zeta$) along $\Gamma-\mathrm{M}$ direction derived from the \textit{d}$_{x^2-y^2}$ orbital, Fig. \ref{figure4}(g). The VBs show linear dispersing bands along the R-Z-A path with different $d$ orbital characters, whose cuts at the E$_\mathrm{F}$ are highlighted by arrows in Fig. \ref{figure4}(j) and (k), again well captured by our DFT calculations. Finally, the flat-\textit{ish} band along the Z-A path is derived also from the \textit{d}$_{xy}$ and \textit{d}$_{z^2}$ orbitals of Co, which contribute 0.62 and 0.22 $\mathrm{\mu}_B$, respectively, to the total magnetization, see suppl. inf. Table I. The detailed electronic structure show small contributions of the pnictide orbitals, with the 3\textit{p} electrons of P and the 4\textit{p} electrons of As contributing to the \textit{p}-\textit{d} bonding at higher binding energies.

We have carried out a comprehensive study of the electronic band structure of CaCo$_2$As$_2$ and LaCo$_2$P$_2$, both from ARPES and DFT. We have demonstrated that the Fermi surface of CaCo$_2$As$_2$ is markedly 3D with dispersive electronic bands in the \textit{k}$_x$-\textit{k}$_z$ plane, but dispersionless in LaCo$_2$P$_2$. Moreover, the experimental Fermi pocket sizes are similar to the DFT predictions, hence indicating that the electronic correlations in (Ca,La)Co$_2$(As,P)$_2$ are weaker than in their sibling iron pnictides. It follows that the dimensionality of the band structure of Co-pnictides is directly linked to the magnetic order and, hence, a consequence of X-X (As-As or P-P) separation between CoX$_2$ planes. The shorter As-As distance in CaCo$_2$As$_2$ ($\simeq$2.80 \r{A}) favors a strong covalent bond along the \textit{c}-direction, featuring a \textit{c}T phase that enhances the magnetic exchange interaction along the \textit{c}-axis that promotes an AFM order, given that the energy difference between the AFM and FM states in CaCo$_2$As$_2$ is rather small. On the other hand, the P-P distance increases by 16\% in LaCo$_2$P$_2$, reducing the coupling between the Co-P planes and forcing an FM order, assisted by the strong in-plane magnetic anisotropy of Co atoms. 

As sibling compounds of superconducting iron pnictides, CaCo$_2$As$_2$ and LaCo$_2$P$_2$ present very different band structures. (Ca,Ba)Fe$_2$As$_2$ exhibits a second-order-like stripe pattern followed by a first-order structural transition to a \textit{c}T phase at low temperature that underscores the emergence of a broken rotational symmetry state \cite{Fernandes2022}. From the electronic structure point of view, the iron pnictides have several hole pockets at $\Gamma$ that nest the electron pockets around the M point, driving the AFM order with a propagation vector ($\pi$-0). As a direct consequence of adding one more electron from Co, the upward shift of the chemical potential by $\sim$1 eV \cite{Ali2017} and the presence of electron pockets away from the X position in CaCo$_2$As$_2$ prevent the in-plane Fermi surface nesting. Interestingly, the ARPES data we report here show parallels to the dimensionality crossover observed in the AFe$_2$As$_2$ family. In particular, a 3D \textit{k}$_z$ dispersion of the electronic band structure develops in the frustrated AFM CaCo$_2$As$_2$, akin to the three-dimensional topology crossover observed in the orthorhombic phase at low temperature in CaFe$_2$As$_2$ \cite{Liu_2009}, yet the spin polarization changes from out-of-plane to in-plane.
Nevertheless, the \textit{c}T phase of Fe pnictides is non-magnetic as a consequence of a major topological change of the Fermi surface that reduces the nesting between the $\Gamma$ and M pockets. Finally, the FM order in LaCo$_2$P$_2$ reconstructs the electronic structure and the number of dispersive bands at low temperature. This is attributed to spin splitting caused by the Stoner-driven magnetic ordering \cite{kovnir2010tuning}, similar to the electronic structure of RbCo$_2$Se$_2$ that shares the same lattice structure \cite{Huang_2021}. 

In summary, we have presented a comprehensive analysis of the band structure that demonstrates a correlation between magnetic order, chemical pressure, and electronic dimensionality in the Co-pnictides CaCo$_2$As$_2$ and LaCo$_2$P$_2$. We have reported a pronounced 2-dimensionality of the electronic structure of LaCo$_2$P$_2$, in contrast to the 3D character of CaCo$_2$As$_2$. The experimental and calculated band structure indicate that these pnictides undergo a large Fermi-surface reconstruction at the magnetic transitions, which are a consequence of the flat-\textit{ish} bands of Co. Moreover, we have shown that the shorter As-As in CaCo$_2$As$_2$, as compaied to the P-P distance in LaCo$_2$P$_2$, enhances electron hopping between the
pnictogen and transition metal sites and, hence, favors the AFM ordering in CaCo$_2$As$_2$. Our findings deepen the fundamental understanding of the electronic band structure of a large family of compounds that host a variety of collective ground states and provide evidence that the structural, electronic, and magnetic properties are strongly entangled.

\section*{Acknowledgments}
D.S., A.K. and S.B-C. acknowledge financial support from the MINECO of Spain through the project PID2021-122609NB-C21 and by MCIN and by the European Union Next Generation EU/PRTR-C17.I1, as well as by IKUR Strategy under the collaboration agreement between IKERBASQUE Foundation and DIPC on behalf of the Department of Education of the Basque Government. C-Y.L. was supported by the European Research Council (ERC) under the European Union’s Horizon 2020 research and innovation program (Grant Agreement No. 101020833). M.G.D. acknowledges financial support from the Government of the Basque Country through the pre-doctoral fellowship PRE 2023 2 0024. M.S. acknowledges the support from the National Science Foundation (award DMR-2233902). This research used resources of the ESM beamline  of the National Synchrotron Light Source II, a U.S. Department of Energy (DOE) Office of Science User Facility operated for the DOE Office of Science by Brookhaven National Laboratory under Contract No. DE-SC0012704. LOREA beamline is co-funded by the European Regional Development Fund (ERDF) within the ”Framework of the Smart Growth Operative Programme 2014-2020”. We acknowledge the MAX IV Laboratory for beamtime on the Bloch beamline under proposal 20210823. Research conducted at MAX IV, a Swedish national user facility, is supported by Vetenskapsrådet (Swedish Research Council, VR) under contract 2018-07152, Vinnova (Swedish Governmental Agency for Innovation Systems) under contract
2018-04969 and Formas under contract 2019-02496. P.G. acknowledges financial support from the MINECO project PID2023-149494NB-C32. M.G.V. thanks support to the
Spanish Ministerio de Ciencia e Innovación (PID2022-142008NB-I00), the Canada Excellence Research Chairs Program for Topological Quantum Matter, and the Diputación Foral de Gipuzkoa Programa Mujeres y Ciencia.
 
\bibliography{bibliography}
\end{document}



\title{Supplementary Material of: `Flat band driven itinerant magnetism in the Co-pnictides CaCo$_2$As$_2$ and LaCo$_2$P$_2$'}

\author{D. Subires}
\email{david.subires@dipc.org}
\affiliation{Donostia International Physics Center (DIPC), Paseo Manuel de Lardizábal. 20018, San Sebastián, Spain}
\affiliation{Applied Physics Department, University of the Basque Country (UPV/EHU), Basque Country, Bilbao, 48080 Spain}

\author{M. García-Díez}
\affiliation{Donostia International Physics Center (DIPC), Paseo Manuel de Lardizábal. 20018, San Sebastián, Spain}
\affiliation{Physics Department, University of the Basque Country (UPV/EHU), Leioa, Spain}

\author{A. Kar}
\affiliation{Donostia International Physics Center (DIPC), Paseo Manuel de Lardizábal. 20018, San Sebastián, Spain}

 \author{C.-Y. Lim}
\affiliation{Donostia International Physics Center (DIPC), Paseo Manuel de Lardizábal. 20018, San Sebastián, Spain}

\author{Victoria M. Li}
\affiliation{Department of Chemistry and
Biochemistry, Florida State University, Tallahassee, Florida
32306, United States}

\author{V Yannello}
\affiliation{Department of Chemistry and
Biochemistry, University of Tampa, 401 W. Kennedy Blvd. Tampa, FL 33606}

\author{Dina Carbone}
\affiliation{MAX IV Laboratory, Fotongatan 2, Lund, 225 92, Sweden}

\author{P. Gargiani}
\affiliation{ALBA Synchrotron Light Source, Cerdanyola del Vall\`es, 08290 Barcelona, Catalonia, Spain}

\author{T. Yilmaz}
\affiliation{National Synchrotron Light Source II, Brookhaven National Laboratory, Upton, New York 11973, USA}
\affiliation{Department of Physics, Xiamen University Malaysia, Sepang 43900, Malaysia}

\author{J. Dai}
\affiliation{ALBA Synchrotron Light Source, 08290 Barcelona, Spain}

\author{M. Tallarida}
\affiliation{ALBA Synchrotron Light Source, 08290 Barcelona, Spain}

\author{E. Vescovo}
\affiliation{National Synchrotron Light Source II, Brookhaven National Laboratory, Upton, New York 11973, USA}

\author{M. Shatruk}
\affiliation{Department of Chemistry and
Biochemistry, Florida State University, Tallahassee, Florida
32306, United States}

\author{Maia G. Vergniory}
\affiliation{Donostia International Physics Center (DIPC), Paseo Manuel de Lardizábal. 20018, San Sebastián, Spain}
\affiliation{Département de Physique et Institut Quantique,
Université de Sherbrooke, Sherbrooke, J1K 2R1 Québec, Canada.}

\author{S. Blanco-Canosa}
\email{sblanco@dipc.org}
\affiliation{Donostia International Physics Center (DIPC), Paseo Manuel de Lardizábal. 20018, San Sebastián, Spain}
\affiliation{IKERBASQUE, Basque Foundation for Science, 48013 Bilbao, Spain}
\date{July 2025}

\maketitle

\appendix
\tableofcontents

\section{\label{app:dft}DFT Calculations}

\begin{figure*}[h!]
    \centering
    \includegraphics[width=\linewidth]{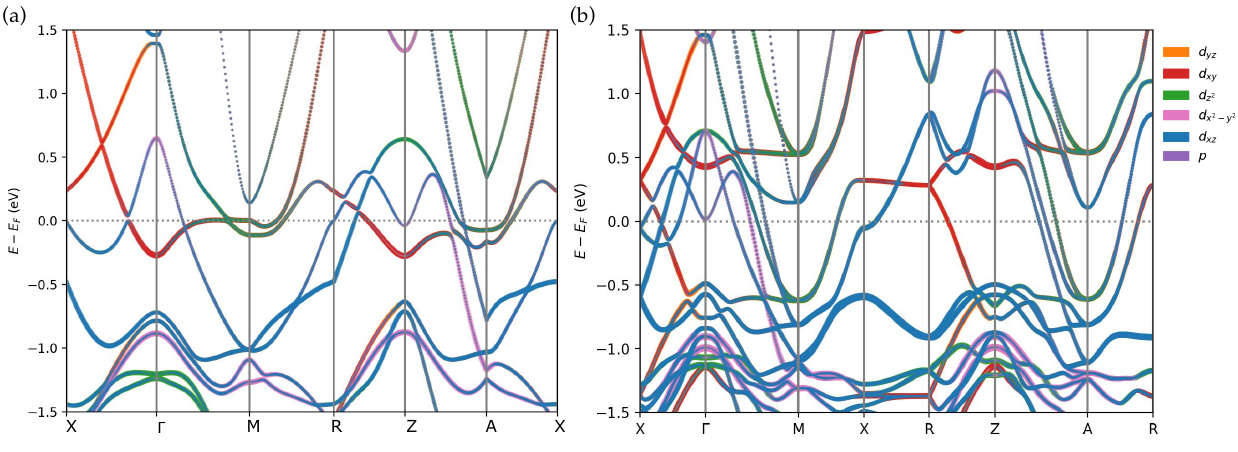}
    \caption{Orbital resolved DFT calculation for $\mathrm{CaCo_2As_2}$ for (a) PM and (b) AFM. The thickness encodes the weight of each atomic orbital component in the Bloch eigenstate.
    }
    \label{figureSupp_CaCoAs_bands}
\end{figure*}
\begin{figure*}[h!]
    \centering
    \includegraphics[width=\linewidth]{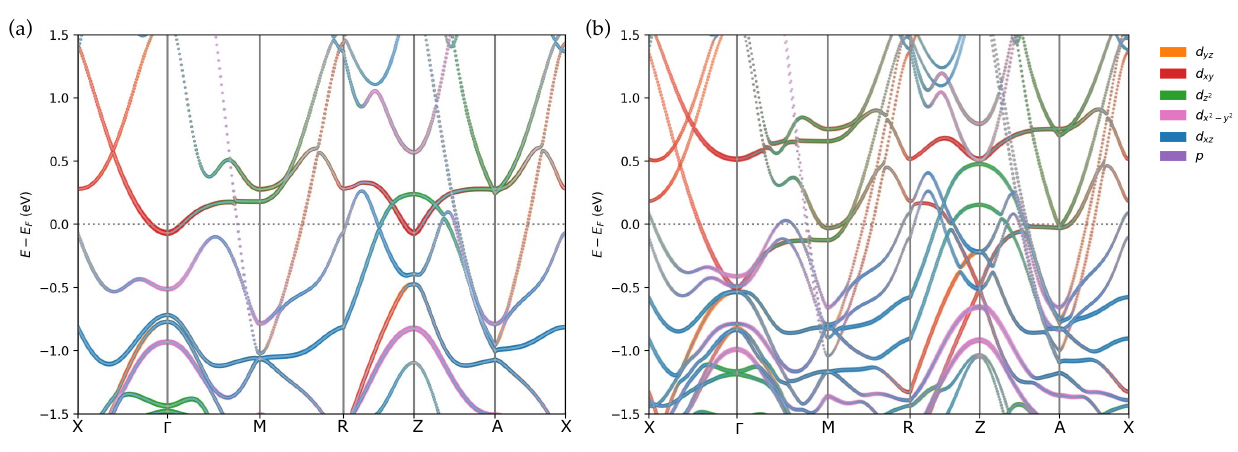}
    \caption{Orbital resolved DFT calculation for $\mathrm{LaCo_2P_2}$ for (a) PM and (b) FM. The thickness encodes the weight of each atomic orbital component in the Bloch eigenstate.
    }
    \label{figureSupp_LaCoP_bands}
\end{figure*}

Figures~\ref{figureSupp_CaCoAs_bands} and~\ref{figureSupp_LaCoP_bands} show the orbital resolved DFT calculations of the bulk bands for the paramagnetic and magnetic ground states of $\mathrm{CaCo_2As_2}$ and $\mathrm{LaCo_2P_2}$, respectively. 

\begin{figure*}
    \centering
    \includegraphics[width=\linewidth]{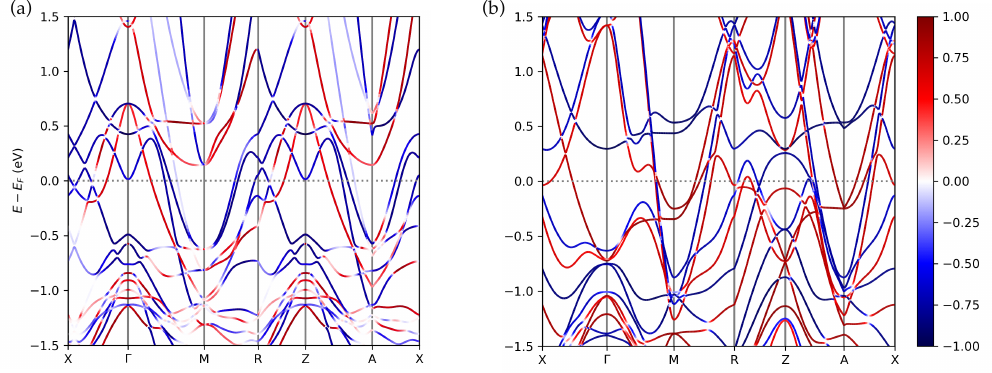}
    \caption{Spin-projected DFT calculations colored according to the projection onto the magnetization axis. (a) $S_z$ projection in AFM $\mathrm{CaCo_2As_2}$ showing that there is no clear separation between spin-up and down bands and Kramers degeneracy. (b) $S_x$ projection in FM $\mathrm{LaCo_2P_2}$ where there is still a good separation of the spin split bands except at points where spin-up and down bands hybridize due to SOC.}
    \label{spin_projection}
\end{figure*}

Figure \ref{spin_projection} presents the spin-projected DFT calculations, highlighting the Kramers degeneracy in CaCo$_2$As$_2$. 

\begin{figure*}[h!]
    \centering
    \includegraphics[width=\linewidth]{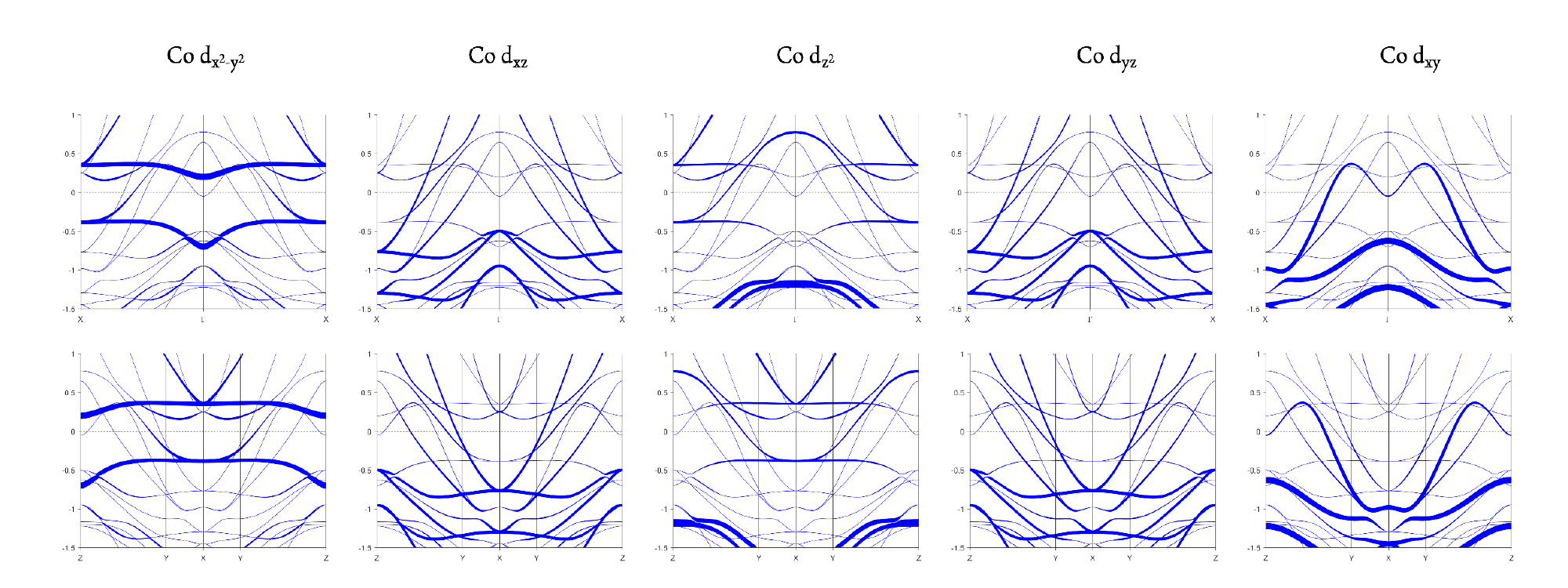}
    \caption{Orbital resolved FM DFT calculation for $\mathrm{CaCo_2As_2}$. Note that the Co band with d$_{xy}$ and d$_{yz}$ character matches the band dispersion of the experimental `$\wedge$' band. 
    }
    \label{figure_FM_CaCoAs}
\end{figure*}

On the other hand, as stated in the main text, the complete and accurate description of the experimental band structure of CaCo$_2$As$_2$ involves a combination of AFM and FM solutions. In figure \ref{figure_FM_CaCoAs}, we show the orbital resolved FM band structure of CaCo$_2$As$_2$, highlighting the good matching of the broad `$\wedge$' band at $\simeq$0.4 eV below the E$_\mathrm{F}$.   

\section{Free energy of CaCo$_2$As$_2$ and LaCo$_2$P$_2$ as a function of the X-X distance}

\begin{figure*}[h!]
    \centering
    \includegraphics[width=0.85\linewidth]{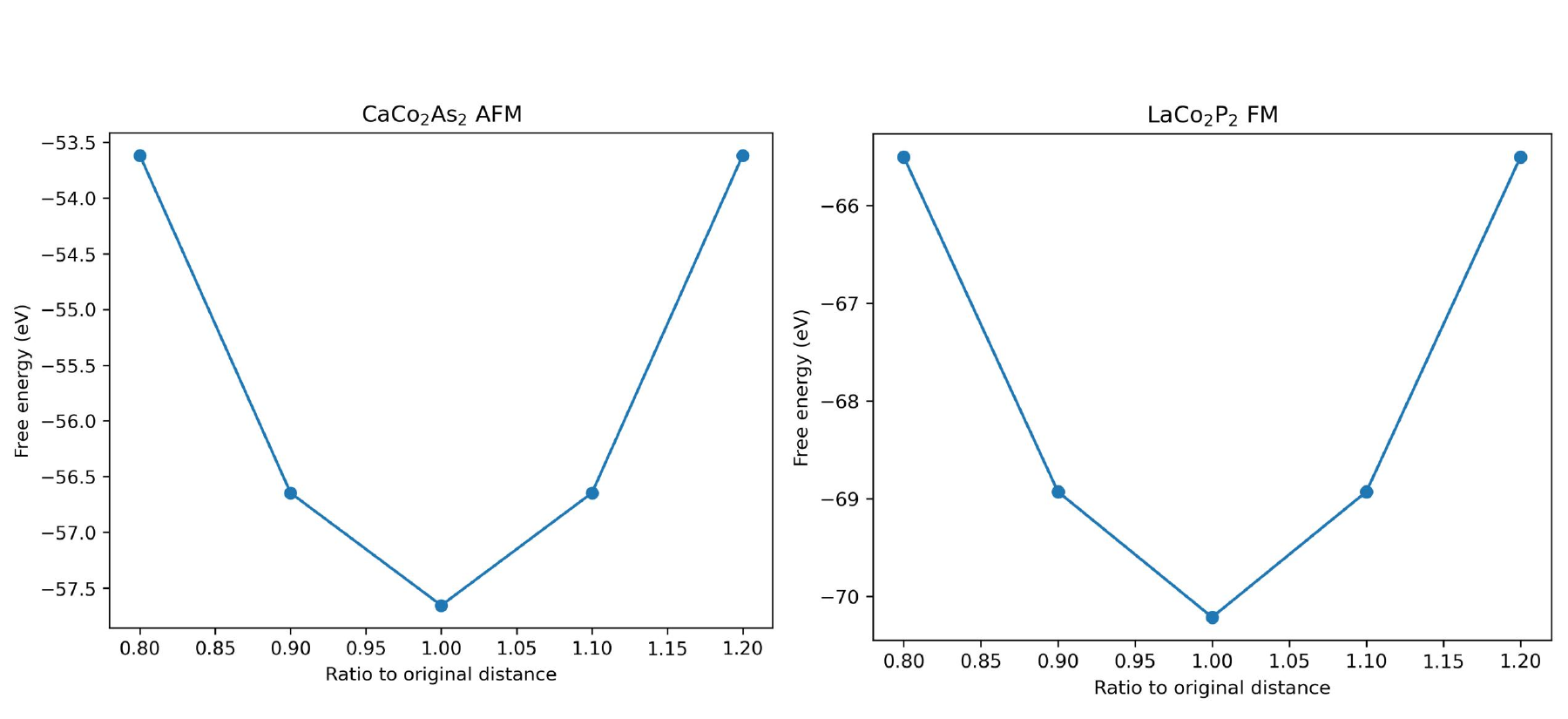}
    \caption{Free energy of the AFM ground state of CaCo$_2$As$_2$ (left panel) and FM LaCo$_2$P$_2$ (right panel) as a function of the X-X distance (As-As and P-P, respectively). The DFT calculations fully agree with the experimental observations.
    }
    \label{X-X}
\end{figure*}

To demonstrate the robustness of the magnetic order, we have calculated the free energy of the ground state of CaCo$_2$As$_2$ and LaCo$_2$P$_2$ as a function of the X-X (As-AS and P-P) distance. As shown in Figure \ref{X-X}, the DFT calculations find that both the AFM and FM ground states are at the minima of the total energy, in agreement with the experimental observations.

\section{\label{app:ARPESCaCo}DFT: surface state calculation of  $\mathbf{CaCo_2As_2}$ and  $\mathbf{LaCo_2P_2}$}

\begin{figure*}[h!]
    \centering
    \includegraphics[width=\linewidth]{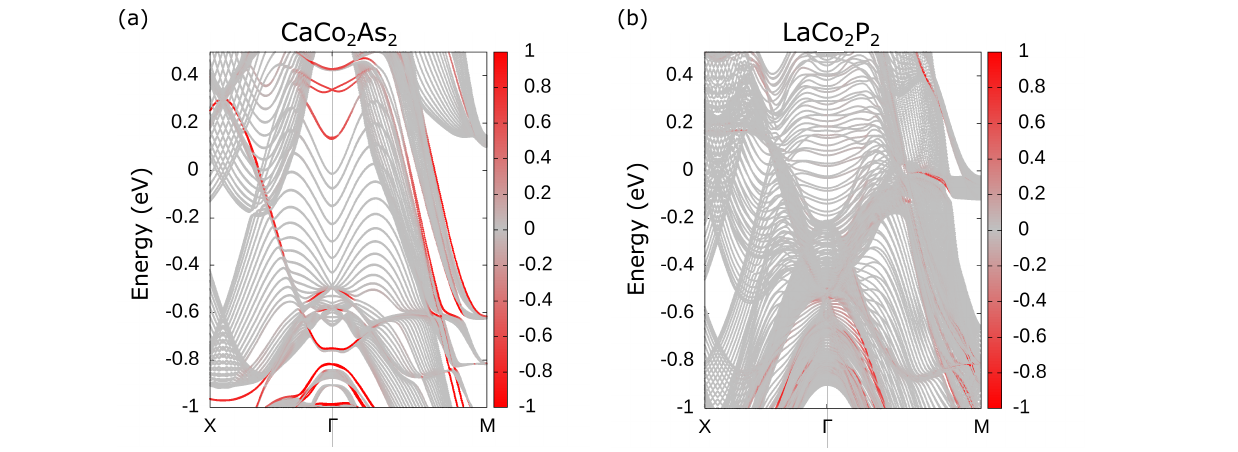}
    \caption{Surface Bands calculation for (a) $\mathrm{CaCo_2As_2}$ and (b) $\mathrm{LaCo_2P_2}$. The bands are computed using a slab with 15 cells in the tetragonal $c$ axis direction using the Wannier tight-binding model. The color represents the weight of the eigenstates on the surface atoms computed from the eigenvectors of the Bloch Hamiltonian.}
    \label{figureSupp_surface_bands}
\end{figure*}

Figure \ref{figureSupp_surface_bands} demonstrates that the experimental ARPES spectra shown in the main text indeed correspond to bulk bands. Both CaCo$_2$As$_2$ and LaCo$_2$P$_2$ surface bands barely contribute to the ARPES spectral weight.  

\begin{table}[h]
    \centering
    \begin{tabular}{ c | c | c  } 
    \hline\hline
        Orbital & $\mathrm{CaCo_2As_2}$ ($\mu_B$) & $\mathrm{LaCo_2P_2}$ ($\mu_B$)\\
        \hline
        $s$     & 0.00872 & 0.01393  \\
        \hline
        $p_y$    & -0.00384 & -0.00422 \\
        \hline
        $p_z$    &  0.00231 & 0.00264  \\
        \hline
        $p_x$    &  -0.00143 & -0.00422 \\
        \hline
        $d_{xy}$   & 0.39144 & 0.62370  \\
        \hline
        $d_{yz}$   & 0.03668 & 0.06977  \\
        \hline
        $d_{z^2}$   & 0.12087 & 0.22284  \\
        \hline
        $d_{xz}$   & 0.03263 & 0.06957  \\
        \hline
        $d_{x^2-y^2}$ & 0.00066 & 0.01668  \\
        \hline\hline
    \end{tabular}
    \caption{Contribution to the final magnetic momenta of the Co atoms separated by orbital. The values are computed from the expectation value of the $\sigma_i$ (where $i=z,x$ for $\mathrm{CaCo_2As_2}$ and $\mathrm{LaCo_2P_2}$, respectively) operators of the Kohn-Sham eigenstates projected onto the spherical harmonics and integrated up to the Fermi level.}
    \label{tab:orbital_values}
\end{table}

\section{ARPES}

\begin{figure*}
    \centering
    \includegraphics[width=\linewidth]{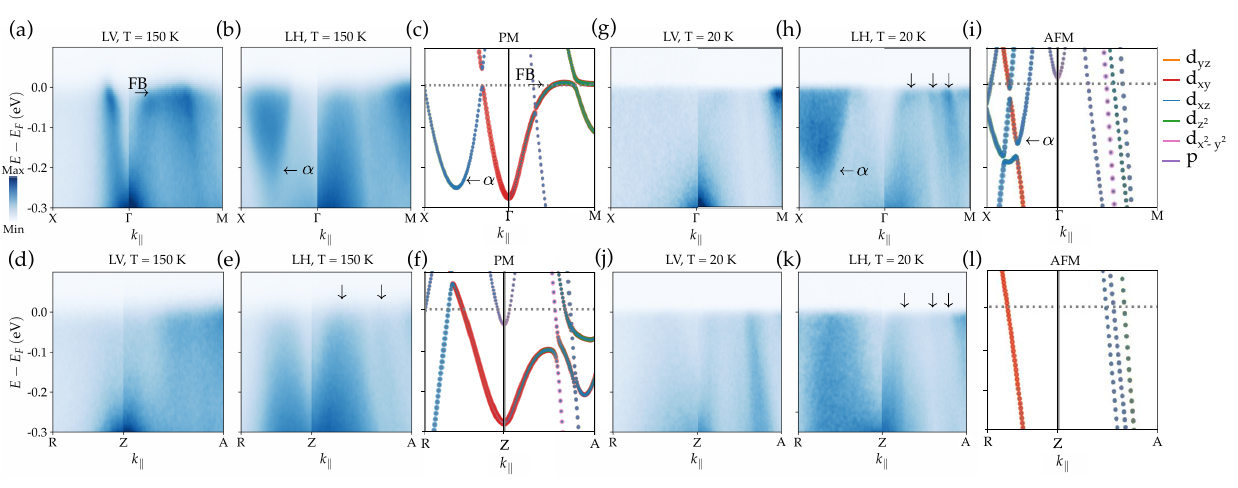}
        \caption{Zoom panels of Figure 3 with binding energy axis $E-E_F = [0.3,\,0.1]\,\mathrm{eV}$. (a-b) Electronic band dispersion of CaCo$_2$As$_2$ along the high symmetry lines $\mathrm{X}-\Gamma-\mathrm{M}$ taken with linear vertical (LV) and horizontal (LH) light, respectively, at $k_z=0$ (E$_\mathrm{i}$=78 eV), T=150 K. (c) Atomic orbital contribution DFT calculation of Co the \textit{d} orbitals in the paramagnetic state at  $k_z\simeq0$. (d,e) Electronic energy bands along the high symmetry lines $\mathrm{R}-\mathrm{Z}-\mathrm{A}$, E$_\mathrm{i}$=105 eV. (f) Atomic orbital contribution DFT calculation of Co the $d$ orbitals at \textit{k}$_z$=$\pi$. (g-h) Band dispersion at low temperature, T=10 K and $k_z\simeq0$. (i) Atomic orbital contribution DFT calculation, $k_z=0$. (j-k) Band dispersion and (l) DFT calculation at $k_z=\pi$.
        }
    \label{figureSupp_CaCoAs_Zoom}
\end{figure*}

\begin{figure*}
    \centering
    \includegraphics[width=\linewidth]{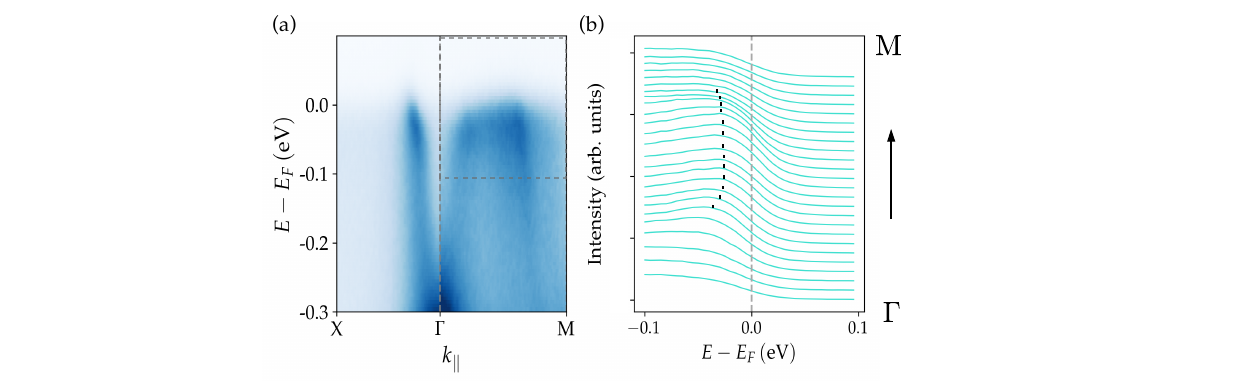}
    \caption{(a) ARPES measurement taken with LV at T = 150 K $\mathrm{CaCo_2As_2}$. (b) EDC analysis of the square region of (a), highlighting the maximum of each EDC.
    }
    \label{figureSupp_CaCoAs_EDC}
\end{figure*}

Figure~\ref{figureSupp_CaCoAs_Zoom} shows a zoom in of the panels of Figure 3 of the main text with higher resolution. In the main text, we discuss the presence of a flat band along the direction $\Gamma-\mathrm{M}$ in $\mathrm{CaCo_2As_2}$ for the paramagnetic phase (see Figure~\ref{figureSupp_CaCoAs_bands} (a)). Figure~\ref{figureSupp_CaCoAs_EDC} shows the ARPES measurement taken with LV light and its corresponding Energy Distribution Curves (EDC) analysis. Panel (b) highlights the presence of a flat band in the EDC spectra.

\clearpage
\bibliographystyle{apsrev4-1}
\bibliography{bibliography}